# High brightness diode-pumped organic solid-state laser

Zhuang Zhao,[1, 2] Oussama Mhibik,[1,2] Malik Nafa,[1,2] Sébastien Chénais[1,2] and Sébastien Forget[1,2 a)]

Author to whom correspondence should be addressed. Electronic mail: sebastien.forget@univ-paris13.fr.

[1] *Université Paris 13, Sorbonne Paris Cité, Laboratoire de Physique des Lasers, F-93430, Villetaneuse, France*

[2] *CNRS, UMR 7538, LPL, F-93430, Villetaneuse, France*

***Abstract :*** *High-power, diffraction-limited organic solid-state laser operation has been achieved in a vertical external cavity surface-emitting organic laser (VECSOL), pumped by a low-cost compact blue laser diode. The diode-pumped VECSOLs were demonstrated with various dyes in a polymer matrix, leading to laser emissions from 540 nm to 660 nm. Optimization of both the pump pulse duration and output coupling leads to a pump slope efficiency of 11% for a DCM based VECSOLs. We report output pulse energy up to 280 nJ with 100 ns long pump pulses, leading to a peak power of 3.5 W in a circularly symmetric, diffraction-limited beam.*

Thin film organic solid-state lasers (OSSLs) have emerged as an important category of coherent sources, offering wavelength tunability over the whole visible spectrum.[1,2] One of the main advantages of organic materials as gain media for lasers is their ability to be easily processed and their intrinsic low cost, making them interesting light sources for many applications such as spectroscopy,[3] bio/chemo-sensing,[4, 5] and polymer fiber telecommunications.[6] However, as electrical pumping is still an open challenge, OSSLs are generally optically pumped by frequency-doubled or tripled solid-state lasers, which are bulky and much more expensive than the organic laser structure itself, jeopardizing the promises of a truly low-cost system. Within the last decade, the spectacular progress in the field of high power inorganic blue laser diodes (LD) and light-emitting diodes (LED) enabled several teams to report on diode-pumped and LED-pumped OSSLs, thus demonstrating true low-cost systems.[7-12] Nevertheless, as the peak power of laser diodes (and all the more LEDs) is weak compared to the values achieved with pulsed solid-state lasers, very low threshold laser resonators are needed. Diode pumping has been achieved up to now through the use of distributed feedback (DFB)[7-9, 12] or organic VCSEL (1D microcavity)[10,11] resonators.

As low threshold operation involves a low output coupling, minimizing threshold and maximizing output power are generally incompatible in a single laser device.[13] As a consequence, the reported output powers of such diode-pumped lasers are very low (either given in arbitrary units [8,9] or when measured, in the pJ [10] or nJ [9] range) and the only reported optical-to-optical efficiencies are around 1%.[9,10] Furthermore, DFB low loss compact resonators do not generally provide diffraction-limited beams,[12] which combined with low output powers leads to a weak brightness, which is a strong limitation especially for applications requiring a free-propagating beam, such as *e.g.* Raman spectroscopy[3]. Brightness is related to spatial coherence and is a figure that is typically orders-of-magnitude higher in lasers than in any incoherent source. The power (or energy, respectively) brightness or radiance is the peak power (energy, resp.) over the optical *étendue*, and is an indication of how much a given optical power (or energy) can be channeled into a given mode or focused to a tight spot.

In this letter, we report on the first diode-pumped vertical external cavity surface-emitting organic laser (VECSOL). This kind of structure[15] (see figure 1 ) has been demonstrated previously under solid-state laser pumping, and enabled obtaining a diffraction-limited output, a high optical-to-optical efficiency (around 50 % [14, 16]), and output energies up to several µJ [15], that is several orders of magnitude higher than what is usually obtained in OSSLs with DFB or microcavity resonators. This is because the output coupling is set to optimize the efficiency (and not minimize the threshold) and because mode area can be made large while keeping the beam quality to the diffraction limit. Reaching threshold in VECSOLs is thus more challenging than in low-loss resonators, but it can now be overcome thanks to the recent availability of high power blue laser diodes (emitting several watts in CW regime) at very reasonable costs.

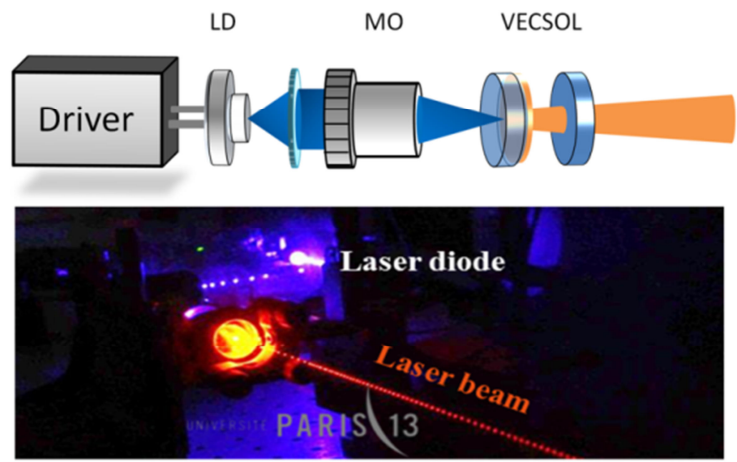

*Fig.1: Schematic diagram of the diode-pumped plano-concave VECSOL resonator (MO : microscope objective, LD : Laser Diode). Inset: Long-exposure time photo of the working diode-pumped VECSOL.*

The VECSOL open cavity enables pump-to-cavity mode matching which is the key for high conversion efficiency, but also offers flexibility for output coupling control and addition of intracavity elements (*e.g.* for wavelength tuning). We studied the influence of the pump pulse duration and output coupler transmission on the laser performance and we achieved a best performance of 3.5 W output peak power. This corresponds to a peak power brightness of 8.9 TW/m$^2$/sr (at a wavelength around 620 nm).

The VECSOL design is shown in figure 1. The gain chip consists of an 18-µm-thick film of dye-doped Polymethylmethacrylate (PMMA, Mw=9.5×10$^5$ g•mol$^{-1}$), spin coated directly onto a highly-reflective plane dielectric mirror (R> 99.5% in the range of 500-670 nm), that is also designed to be transparent (T>94%) at the pump wavelength of 450 nm. The remote output coupler has a radius-of-curvature of 200 mm and is set 1 mm away from the active mirror.

The pump laser source is a low-cost high power blue InGaN laser diode (maximum output power of 1.7 W, central emission wavelength 450 nm) driven by a pulsed power supply (Picolas GmbH), with controllable pulsewidth. The pump beam is focused to a 90 µm*100 µm -in-diameter spot size onto the dye-doped PMMA layer through a high-numerical aperture long-working-distance microscope objective, reasonably matched to the cavity mode (130 µm in diameter).

The time evolution of organic laser emission is recorded with a reverse biased Si pin photodiode attached to a 1 GHz oscilloscope (Agilent) and is synchronized to the pump pulses. All the fabrication process and measurements are made in ambient air with no specific encapsulation of the samples.

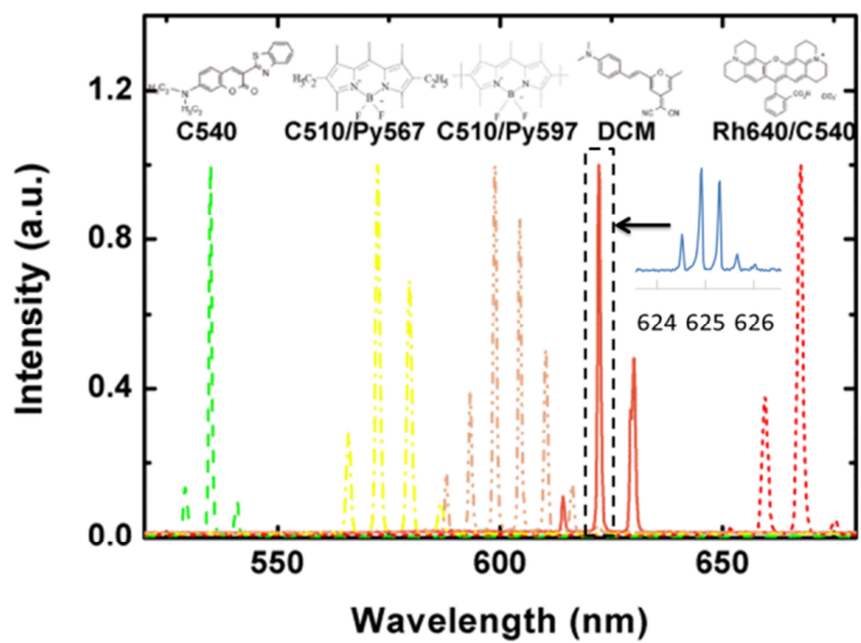

*Fig. 2 : Laser spectra measured from LD pumped VECSOLs with five different dyes, from left to right: C540 (green dash), Py567/C510 (yellow dash-dot), Py597/C510 (light orange dash-double-dot), DCM(orange solid line) and Rh640/C540 (red dot). Inset : high-resolution spectrum (0.1 nm) showing the structure of each peak (here for a 450 µm long cavity).*

To illustrate the versatility of organic laser, we used in this work different lasing dyes to cover a large part of the visible spectrum: Coumarin 510 (C510), Coumarin 540 (C540), Pyrromethene 567 (Py567), Pyrromethene 597 (Py597), 4-Dicyanomethylene-2-methyl-6-p dimethylaminostyryl-4H-pyran (DCM) and Rhodamine 640 (Rh640). All active materials were dispersed into PMMA either alone or in the form of donor/acceptor mixtures (host-guest blends) according to their absorption properties at the pump wavelength. For C540 and DCM, the weight ratio in PMMA is set to 1% which enables the absorption at 450 nm to be 80% in a single pass, while circumventing concentrating quenching. For Py567, Py597 and Rh640, direct pumping at 450 nm is not efficient because of a too low absorption cross-section at this wavelength. We used C510/Py567, C510/Py597 and C540/Rh640 as excitation donor/acceptor pairs, in which the donor dye absorbed the pump light and transferred its excitation to the acceptor dye by a Forster resonant energy transfer (FRET) mechanism.[17] In those cases we blended the "donor" and the "acceptor" with the same weight ratio of 1% in PMMA. Figure 2 shows the lasing spectra recorded using an optical spectrometer (USB 2000+, Ocean optics) with a resolution of ~ 0.3 nm. The evenly spaced peaks in the spectrum correspond to the free spectral range of the Fabry-Perot etalon formed by the 18µm-thick active layer itself. Each of these peaks contains several thin peaks (linewidth below 0.1 nm, limited by our spectrometer resolution) corresponding to the modes of the cavity (see inset in figure 2). Inserting an additional Fabry-Perot filter allowed single-peak operation[18] whereas continuous tunability over 10 nanometers could be achieved upon taking advantage of the thickness variation at the edge of the sample.[15]

In the following, the laser characterization and optimization are detailed with the DCM-based VECSOL.

Beside their interest as low cost pumping sources, diode lasers offer the advantage that pulse duration is totally user-controllable from a few ns to the CW regime. This is of utmost interest for studying organic lasers, where the laser dynamics is complex and strongly linked to the long-lived triplet state population, causing triplet-triplet absorption and triplet-singlet quenching that affect the maximum pulsewidth obtainable from an OSSL. [19]

Figure 3 shows measurements of the VECSOL temporal dynamics when operating with pump pulse widths of 50 ns, 100 ns, 200 ns and 500 ns. The clearly visible delay between the onsets of pump and laser pulses is closely related to the oscillation build up time and therefore depends on the cavity length, the transmission of the output coupler and the pump power density. The clearly visible delay between the onsets of pump and laser pulses is closely related to the oscillation build up time and therefore depends on the cavity length, the transmission of the output coupler and the pump power density. This large delay can be well accounted by solving Statz-DeMars coupled population equations, as detailed in ref 16 in a slightly different case under laser pumping, and is a salient feature of VECSOLs, due to the fact that the active medium thickness represents only a tiny fraction of the cavity length. In figure 3 (a-d), pump and laser pulses are displayed for various pump durations. A typical relaxation oscillation is observed at the beginning of the laser pulse. For long pump pulses (fig. 3 (b-d)), the laser intensity vanishes during the pump pulse, as a consequence of the slow increase of the triplet state population. Those results show that the maximum pulsewidth of a DCM-based VECSOLs is around 100 ns (FWHM) and is limited by triplet population buildup, as also observed in ref. 19.

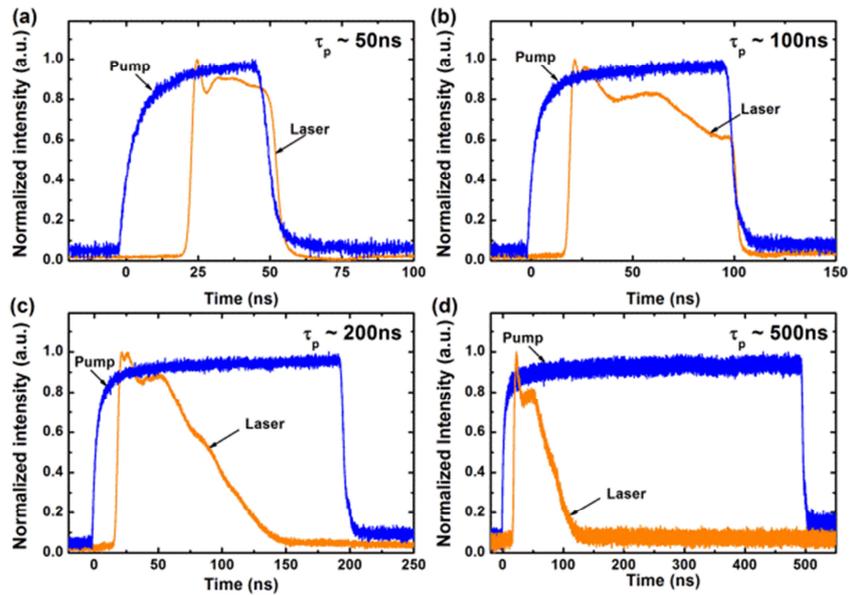

*Fig. 3 Temporal profile of 1%-DCM VECSOL (orange line) pumped with varied pump pulse durations (blue line): (a) 50 ns; (b) 100 ns; (c) 200 ns; (d) 500ns. An output coupler with a transmission of 3% and a peak pump power density of 132 kW/cm$^2$ were used in the experiment.*

The variation of the laser slope efficiency (slope of the laser pulse energy versus the incident pump pulse energy curve) with pump pulse duration is plotted in figure 4. When a 20-ns pump pulse is used, the efficiency is only 2.5%. as the laser pulse starts only at the very end of the pump pulse. The laser slope efficiency increases with the pump pulse width and reaches a maximum value of 11% (corresponding to a quantum slope efficiency of ~ 16%) with a pump pulse width of 100 ns, and then decays due to the triplet population buildup.

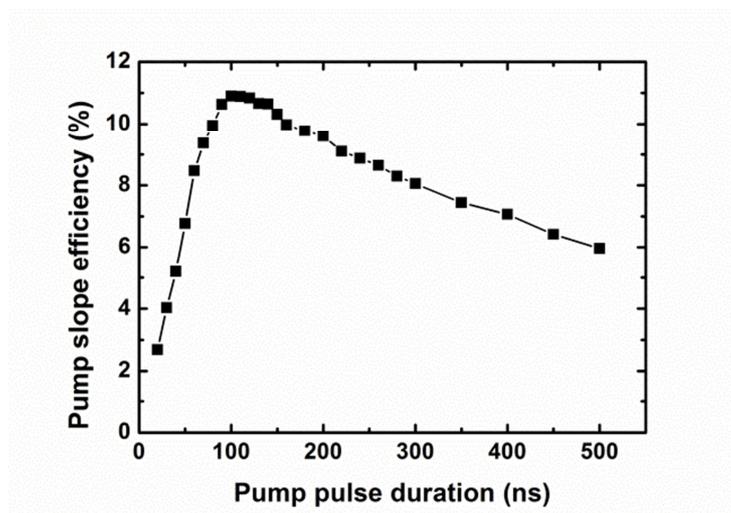

*Fig. 4 pump slope efficiency of 1%-DCM VECSOLs at varied pump pulse duration from 20 ns to 500 ns. The maximum pump slope efficiency was obtained with a pump pulse duration of ~ 100 ns.*

Another factor playing a key role for power extraction in lasers is the choice of the transmission of the output coupler. [13] Whereas every laser cavity has an optimal output coupler transmission (resulting from the trade-off between cavity photon lifetime and energy extraction from the resonator), this feature is not easily controlled in typical monolithic 2$^{nd}$ order DFB structures where the output laser beam comes from the diffraction of the laser wave on the grating itself. In a VECSOL structure, changing the amount of output coupling is straightforward and can be done without affecting the gain properties, providing an additional degree of freedom to optimize the resonator. To this end, the input-output characteristic curves of a 1-mm long VECSOL (with optimal pump-pulse duration of 100 ns, corresponding to a laser pulsewidth of 80 ns) are compared (figure 5) for four different output couplers having transmission of 0.5%, 2%, 3% and 6% respectively. As expected, the minimum pump threshold (32 kW/cm$^2$) is obtained using the output coupler with the lowest transmission (0.5%). The optimum coupler transmission in order to achieve the maximum pump slope efficiency of 11% is 3%; in this case, the pump density threshold is 69 kW/cm$^2$.

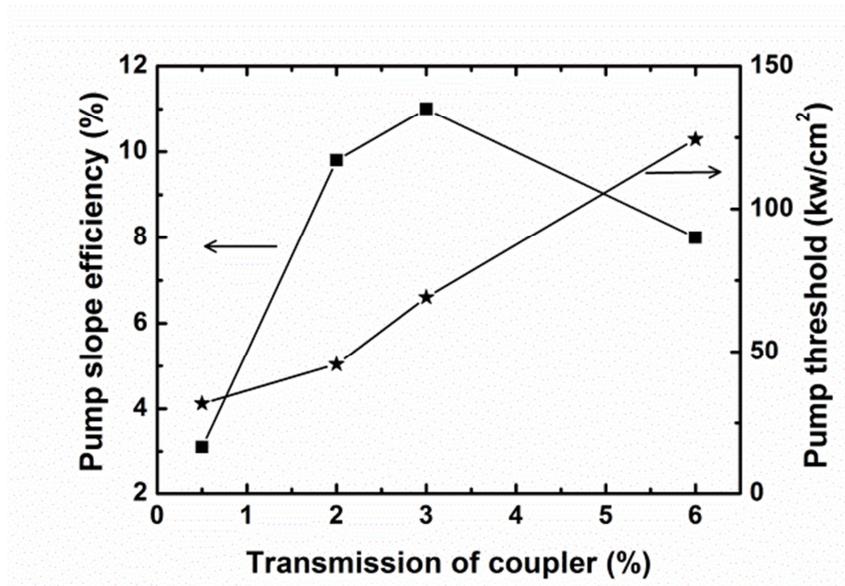

*Fig. 5 Laser performance including pump slope efficiency and pump density threshold, were investigated by changing the transmission of output couplers: 0.5%, 2%, 3% and 6%.*

In figure 6, the emission pulse energy is given as a function of incident pump pulse energy for two output couplers. With a single LD pumping system, the maximum pump pulse energy deposited on the VECSOL chip is 1.6 µJ in a 100 ns pulse, and a maximum emission pulse energy of 130 nJ is achieved with a 2% transmission coupler. Using the 3% transmission coupler, the maximum emission pulse energy is slight lower (115 nJ) while it shows higher pump slope efficiency. To push the output power to its maximum, we drove two LDs with a single LD driver. The two diodes were polarization-coupled through a polarizing beam splitter to provide an incident pump pulse energy on the VECSOL chip of 3.2 µJ in a 100 ns pulse. An energy of 280 nJ is obtained with the 3% transmission coupler, and the peak power (estimated from the pulse energy over FWHM of laser pulse duration) is around 3.5 W. Beam quality was quantified by measuring the M² factor, shown in the inset of figure 6. The LD pumped VECSOL emits a diffraction limited beam with $M^2 \sim 1.02$. A typical image of the $TEM_{00}$ beam profile is also shown. The measured power brightness, defined as $P/(M²)^2 \cdot \lambda^2$, where P is the peak power (in W) and λ the wavelength, is then 9.1 TW/m²/sr at 620 nm. This figure is not easily comparable with other OSSLs, as the M² factor is rarely measured. However, for DFB lasers, its value is always greater that one.[20, 21] M² values range from a few units (an astigmatic annular mode with a M² value as low as 2.2 has been reported in a two dimensional DFB resonator pumped by a solid-state microchip laser,[21]) to several tens (a high divergence of 5° is described for example in ref 22). As a consequence, the highest power brightness (resp. energy brightness) that can be deduced (at 620 nm, with the optimistic hypothesis of a state-of-the art value of 2.2 for M²) from the literature for LD-pumped OSSL is around 54 GW/m²/sr (resp. 1.1 kJ/m²/sr).[9] The herein reported power brightness of 9.1 TW/m²/sr thus corresponds to a two-order of magnitude increase. For the energy brightness, our value of 700 kJ/m²/sr is more than 600 times higher than the previously reported results.

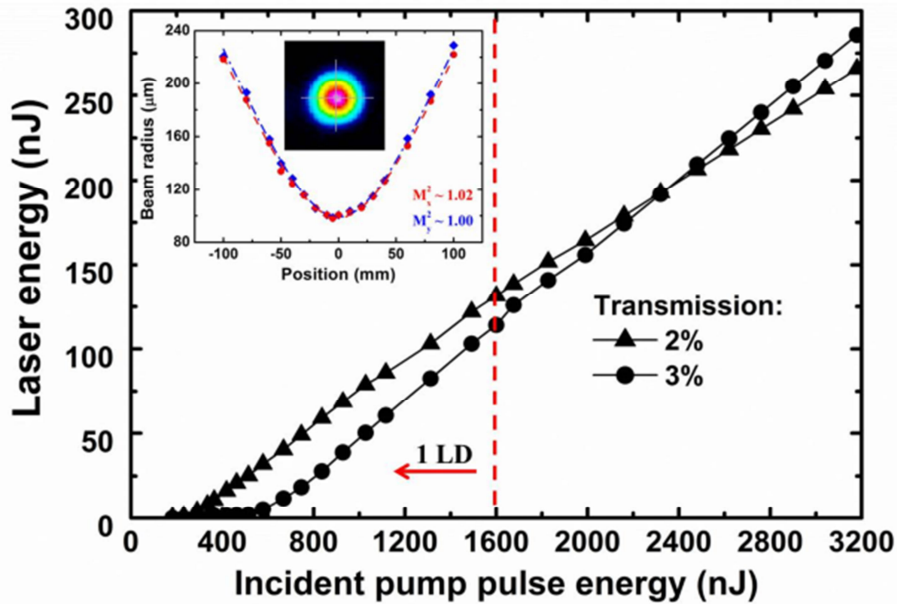

*Fig. 6 Input-output characteristic of 1%-DCM VECSOL, with 1 LD (below 1.6 µJ) and 2 polarization-coupled LDs (above 1.6 µJ.) as the pump. Inset: spatial profile of the output beam and its evolution along the emission direction, yielding $M^2$ ~1 in both transverse directions (diffraction-limited output.)*

Finally, photostability was investigated at fixed pump density above threshold (80 kW/cm²). Under 100-ns long pulse pumping, the output emission decreased to half its initial value after ~ 1000 pulses. As VECSOLs are made of planar, uniform organic layers, it is straightforward to scan the sample (perpendicularly to the cavity axis) to enhance the lifetime. As an example, scanning a 1-inch-in-diameter sample with a 100-µm-in-diameter pump spot, allows more than 8000 fresh spots to be used (with no realignment issue),[18] leading to a lifetime of $8\times10^6$ pulses (equivalent to 13 min operation at 10 kHz). Furthermore, encapsulation can be used since it has shown a significant improvement in the lifetime of the organic device. [1,2]

In summary, we have demonstrated diode-pumped VECSOLs that cover the visible range from 450 nm to 660 nm, by choosing proper dyes or dye mixtures. The lasing performance of diode-pumped VECSOLs was optimized by varying the pump pulse duration and the output coupler transmission. An optimum pump slope efficiency is demonstrated for 100 ns pump pulses, corresponding to a duration long enough to allow for efficient build-up of laser oscillation and short enough to avoid losses and quenching resulting from the buildup of a triplet state population. With DCM-based VECSOLs, a pump slope efficiency of 11% has been obtained with a diode pump pulse width of 100 ns. A maximum energy level of 280 nJ (3.5 W peak power), limited by the available pump pulse peak power (32 W), has been obtained with two polarization-coupled laser diodes, with a nearly diffraction limited beam. Those results pave the way toward the development of truly low-cost, compact, high-brightness tunable organic thin-film solid-state lasers at the µJ level.

**Acknowledgements:**

The authors wish to acknowledge the Agence Nationale de la Recherche (ANR-12-EMMA-0040 "Vecspresso" project) for funding this work.

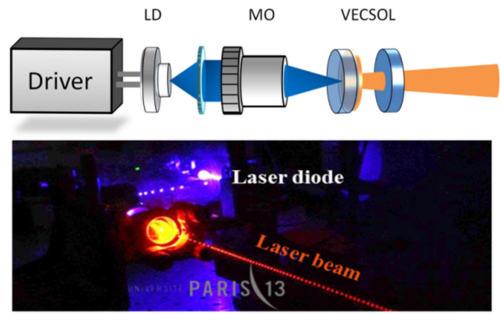

Fig.1: Schematic diagram of the diode-pumped plano-concave VECSOL resonator (MO : microscope objective, LD : Laser Diode). Inset: Long-exposure time photo of the working diode-pumped VECSOL.

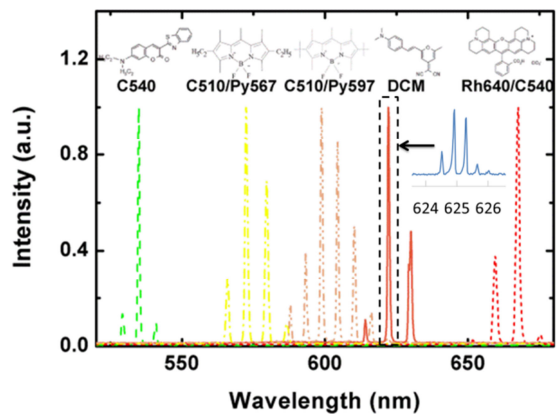

Fig. 2 : Laser spectra measured from LD pumped VECSOLs with five different dyes, from left to right: C540 (green dash), Py567/C510 (yellow dash-dot), Py597/C510 (light orange dash-double-dot), DCM(orange solid line) and Rh640/C540 (red dot). Inset : high-resolution spectrum (0.1 nm) showing the structure of each peak (here for a 450 µm long cavity).

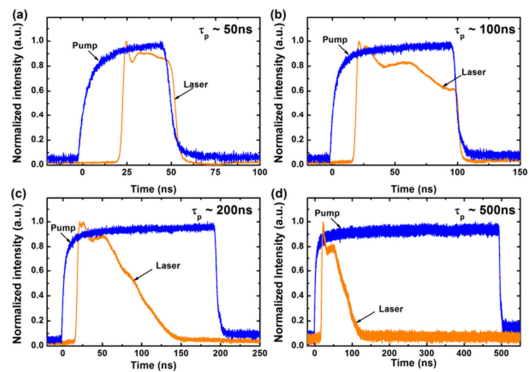

Fig. 3 Temporal profile of 1%-DCM VECSOL (orange line) pumped with varied pump pulse durations (blue line): (a) 50 ns; (b) 100 ns; (c) 200 ns; (d) 500ns. An output coupler with a transmission of 3% and a peak pump power density of 132 kW/cm$^2$ were used in the experiment.

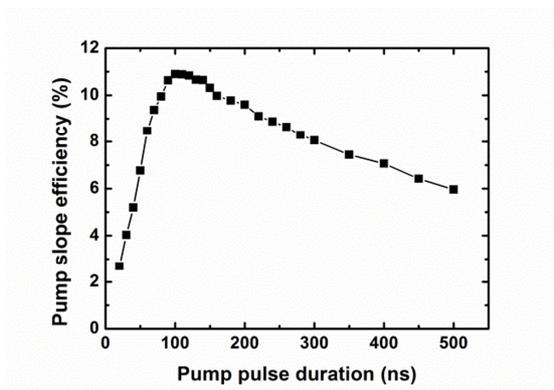

Fig. 4 pump slope efficiency of 1%-DCM VECSOLs at varied pump pulse duration from 20 ns to 500 ns. The maximum pump slope efficiency was obtained with a pump pulse duration of ~ 100 ns.

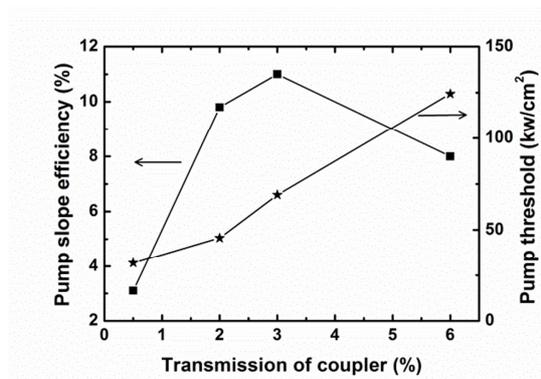

Fig. 5 Laser performance including pump slope efficiency and pump density threshold, were investigated by changing the transmission of output couplers: 0.5%, 2%, 3% and 6%.

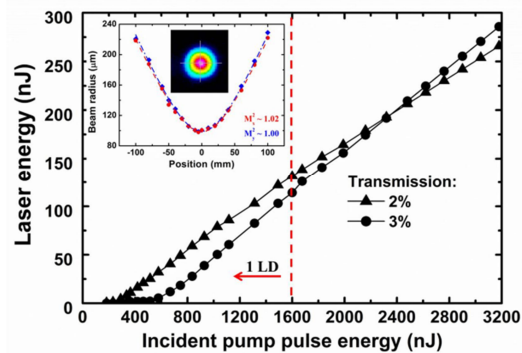

Fig. 6 Input-output characteristic of 1%-DCM VECSOL, with 1 LD (below 1.6 µJ) and 2 polarization-coupled LDs (above 1.6 µJ.) as the pump. Inset: spatial profile of the output beam and its evolution along the emission direction, yielding $M^2$ ~1 in both transverse directions (diffraction-limited output.)